# Are Short and Long Gamma Ray Bursts Really of Different Origin?

Ernst Karl Kunst
Im Spicher Garten 5
53639 Königswinter
Germany
e-mail: ErnstKunst@aol.com

**Short and long gamma ray bursts (GRBs) are of the same origin and, furthermore, correlated with their duration, as will be shown in the following.**

**Key words:** Gamma ray bursts - origin - vacuum Cerencov radiation

According to a NASA press release, Jay Norris of the Goddard Space Flight Center found that the shorter GRBs, lasting less than 2 seconds (s), have different charasteristics than longer bursts [1]. Because I could not find the original paper in the literature I refer in the following to this press release, especially since the general results given there are fully sufficient to be compared with some theoretical derivations.

According to Norris have short GRBs (< 2 s) significantly fewer pulses and are their lag times (the lag of lower-energy pulses behind high-energy pulses) 20 times shorter than the lags in the longer GRBs. Therefore, he proposed that the short bursts are produced in physically different objects.

In the following is demonstrated that these experimental findings do exactly coincide with theory predicting a common origin of all GRBs. In [2] the outlines of this theory have been drawn, according to which GRBs and related phenomena in the X-rays and the ultra-violet have their origin not in known or speculative astronomical objects but rather in the vacuum Cerenkov radiation caused by the superluminal propagation of extraterrestrial spaceprobes in the interstellar space (see also [3]). Cerenkov radiation in all wavebands is generated along the flight paths of the superluminal spaceprobes, whereby the photons of highest energy depend on the superluminal velocity of the probe or craft. On the grounds of the Cerenkov angle and the known duration of GRBs their distance is basically calculable. Furthermore, a correlation between duration, distance, relative number and intensity of GRBs has been shown to exist. It is demonstrated that this correlation also comprises the above stated decrease of pulses and shortening of lag times in short GRBs.

According to theory the Cerenkov radiation generated by a superluminal extraterrestrial spaceprobe hurtling through the galactic space is emitted along and in a very narrow cone in the direction of its flight path. An observer in the vicinity of Earth will observe a GRB exactly then, when the spaceprobe crosses the line of sight in a very narrow angle (Fig. 1) and the generated Cerenkov radiation is bright enough to be observed. The radiation cone comprises - depending on the velocity - photons of all wavebands till down to the radio waveband.
Fig. 1 shows (exaggerated) that the "visible" gamma light track in the sky constitutes the side BC of the triangle ABC, where AB and AC are the "lines of sight" to the points of the track wherefrom photons of lowest (B) and highest (gamma) energy (C), respectively, are received and the distance coincides with AC of this triangle. Due to the different Cerenkov angle of the respecitve radiation will the Cerenkov point source be observed to recede backward in time along BC from point C to point B and beyond with decreasing frequency from ever more distant points with ever lower velocity in the plane of the sky.

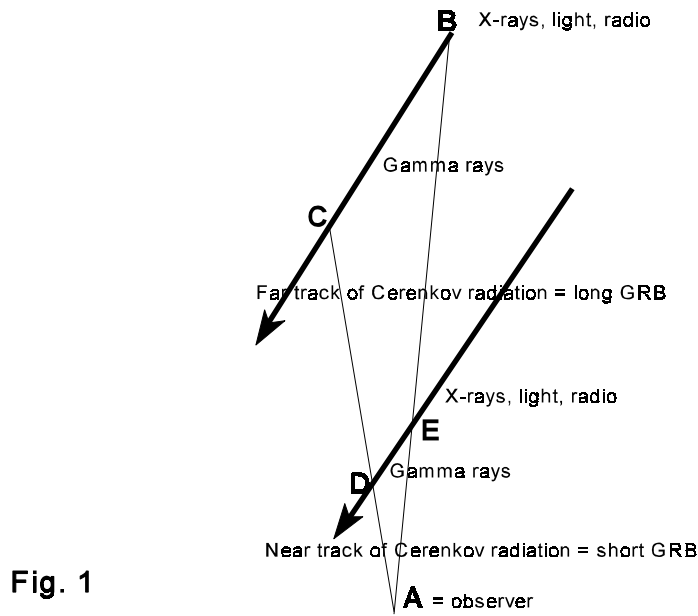

Fig. 1

Fig. 1 shows two tracks of vacuum Cerenkov radiation which we assume be generated by spaceprobes crossing the line of sight of the observer at point A with the same superluminal speed - implying them to be alike in all physical properties - but in different distances. It is clearly evident that the respective distance is proportional to the length of the observable Cerenkov track of gamma radiation BC and DE, respectively, which again is proportional to the duration of the GRB. On the other hand, the tracks BC and DE are, independent on their different length or duration, exactly of the same spectral composition, with the photons of the same highest energy at the points B and D and of lowest energy at the points C and E, respectively. As has extensively been shown in [2] will owing to the track geometry the photons (pulses) of highest energy generally arrive first at the observer in point A and subsequently photons (pulses) of ever lower energy. Therefore, in connection with Fig. 1 is clear that BC and DE exhibit exactly the same energy spectrum, but the latter on a much shorter scale so that firstly the number of photons or pulses must be proportional and secondly lag times or arrival times of photons of different energy inversely proportional to $BC/DE = D_{BC}/D_{DE} \approx n_{BC}/n_{DE}$, where D means duration and n number of bursts in the respective distance or of the respective duration within some time. If a mean of 30 - 40 s for the longer and of 1.5 - 2 s for the short GRBs is assumed this results in $\approx$ 20 - the result of Norris -, exactly as it does if the number of bursts with a duration of $\approx$ 30 s (29) is divided by the number with a duration of 1.5 s (1.45) [4] (the latter number indeed has been calculated, because the mean burst distribution at 1.5 s is disturbed by a small event maximum (see [2]) and, therefore, may not look very convincing).

Furthermore, follows from this hypothesis that the duration times of the subsequent Cerenkov radiation in the X-rays, visible light etc. ("afterglow") of GRBs obey the same law. Therefore, in the case of very short GRBs a shortening of the duration times of this radiation in accord with the above relation is to expect, which presumably is the cause that it has not been detected yet.

<div align="center">References</div>